\documentclass[10pt]{article}
\usepackage{amsfonts,amssymb,amsmath,comment}
\usepackage{graphicx,graphics,color}
\usepackage{caption}
\usepackage{subcaption}
\usepackage{pgf,tikz}
\usetikzlibrary{arrows}

 \makeatletter
 \@addtoreset{equation}{section}
 \makeatother

 \newcounter{extralabel}[section]

 \newtheorem{ittheorem}{Theorem}
 \newtheorem{itlemma}{Lemma}
 \newtheorem{itproposition}{Proposition}
 \newtheorem{itdefinition}{Definition}
 \newtheorem{itcorollary}{Corollary}
 \newtheorem{itconjecture}{Conjecture}
 \newtheorem{itremark}{Remark}

 \newcommand{\eps}{\epsilon}

\newcommand{\La}{\Lambda}
\newcommand{\la}{\lambda}
\newcommand{\ga}{\gamma}
\newcommand{\Ga}{\Gamma}

\newcommand{\und}{\underline}

\newcommand{\TT}{\mathcal T}

\newcommand{\QQ}{\mathcal Q}
\newcommand{\LLd}{\mathcal L^{(2)}}

\newcommand{\RRd}{\mathcal R^{(2)}}

\newcommand{\be}{\begin{equation}}
\newcommand{\ee}{\end{equation}}

 \newenvironment{theorem}{\addtocounter{extralabel}{1}
 \begin{ittheorem}}{\end{ittheorem}}

 \newenvironment{lemma}{\addtocounter{extralabel}{1}
 \begin{itlemma}}{\end{itlemma}}

 \newenvironment{definition}{\addtocounter{extralabel}{1}
 \begin{itdefinition}}{\end{itdefinition}}

 \newenvironment{corollary}{\addtocounter{extralabel}{1}
 \begin{itcorollary}}{\end{itcorollary}}

 \newenvironment{remark}{\addtocounter{extralabel}{1}
 \begin{itremark}}{\end{itremark}}

\newcommand{\halmos}{\rule{1ex}{1.4ex}}

\begin{document}

\title{Phase transitions and coarse-graining for a system of particles in the continuum}

\author{\renewcommand{\thefootnote}{\arabic{footnote}}
Elena Pulvirenti
\footnotemark[1]
\\
\renewcommand{\thefootnote}{\arabic{footnote}}
Dimitrios Tsagkarogiannis
\footnotemark[2]
}

\footnotetext[1]{Mathematical Institute, Leiden University, P.O.\ Box 9512,
2300 RA Leiden, The Netherlands,\\
\emph{pulvirentie@math.leidenuniv.nl}}

\footnotetext[2]{Department of Mathematics, University of Sussex,
Brighton BN1 9QH, UK,\\
\emph{D.Tsagkarogiannis@sussex.ac.uk}}

\date{\today}

\maketitle

\begin{abstract}We revisit the proof of the liquid-vapor phase transition for systems with finite-range interaction by Lebowitz, Mazel and Presutti \cite{LMP} and extend it to the case where we
additionally include a hard-core interaction to the Hamiltonian. 
We establish the phase transition for the mean field limit and then we also prove it when the interaction range is long but finite, by perturbing around
the mean-field theory.
A key step in this procedure is the construction of a density (coarse-grained) model via
cluster expansion. In this note we present the overall result 
but we mainly focus on this last issue.


\medskip\noindent
{\it Acknowledgment.} We would like to thank Errico Presutti for introducing us to this subject.
EP is supported by ERC Advanced Grant 267356-VARIS.

\end{abstract}

 \maketitle


\section{Introduction}
\label{sec:1}

One of the main open problems in equilibrium statistical mechanics is to prove the validity of a
liquid-vapour phase transition in a continuum particle system.
Although this is well observed in experiments as well as in continuum theories, a rigorous
proof for particle systems is still lacking. Intermolecular forces are often described by Lennard-Jones 
interactions, however the difficulty of handling such  (or more realistic) systems  has promoted
the introduction of several simplified models.
A good compromise between realistic models of
fluids and mathematically treatable systems may consist of particles interacting via
a combination of hard spheres (for repulsion) and an attractive long-range Kac interaction.
However, the free energy of hard spheres can be studied for very small values of the density,
far from the value at which a transition occurs. Hence, we still need to use a long range $4$-body repulsive term as in \cite{LMP} to determine the phase transition point.
Then, the hard-core interaction acts just as a perturbation to the mean-field case. In fact, we show that
the liquid-vapour transition persists if the volume of the hard spheres is sufficiently small,
but finite.
Nevertheless, our model presents a richer behaviour and if one manages to deal with a higher density regime,
the hard-core interaction will become relevant and responsible for another transition of the
gas-solid type.

Our proof will follow Pirogov-Sinai theory in the version proposed by Zahradn{\'i}k, \cite{Z}.  
The analysis requires first of all the notions of \emph{coarse-graining} and \emph{contours} which are introduced in Section \ref{sec:con}
and subsequently, with an argument \'a la Peierls, one has to prove that contours are improbable which we do  in Section \ref{sec:main}. 
In this scenario we are able to compute the effective Hamiltonian for the coarse-grained system with a multi-canonical
constraint (given by the fixed density in each cell). This
computation involves an integration over the positions of the particles in each cell leading to a new
measure on the density at the cells.
The computations which lead to the effective Hamiltonian are in general very complicated, 
nevertheless due to the choice of the
interaction they can be carried out. The crucial point here is to show convergence 
of  a cluster expansion
in the canonical ensemble with hard-core, Kac  interaction and contour weights. 
 This is done in Section \ref{sec:clu} by extending the results in \cite{PT}.

For more details on the proofs we refer to \cite{P}, from which the present paper is a follow-up, to \cite{PT1} and to the monograph
of Presutti \cite{Pre}. 

\section{Model}
\label{sec:2}
We consider a system of identical point particles in $\mathbb{R}^d$, $d \geq 2$, and call \emph{particle configuration} a countable,
locally finite collection of points in $\mathbb{R}^d$. 
The \emph{phase space} $\QQ^{\La}$ is the collection of all particle configurations in a bounded region $\La$. We use the notation $\QQ$ 
when $\La\equiv \mathbb{R}^d$. 
We write $q=(q_1,...q_n)$ to indicate a configuration of $n$ particles positioned at points $q_1,...,q_n$ (the order is not important) 
of $\mathbb{R}^d$, while we write  $q_{\La}$ when we want to specify that the particles are in $\QQ^{\La}$.

We consider a mean field model with an energy density given by:
\be\label{mfield}
e_{\la}(\rho)= -\la\rho- \frac{\rho^2}{2} + \frac{\rho^4}{4!},
\ee
where $\lambda$ is the chemical potential.
Here, the density $\rho=n/|\La|$ is set equal to the total density  and it is therefore constant.
We further define the LMP model, \cite{LMP}, by relaxing
to a local mean field: the Hamiltonian (for configurations with finitely many particles) is given by the following function
\be
H^{\text{LMP}}_{\la,\ga}(q)=\int_{\mathbb{R}^d} e_{\la}(\rho_{\ga}(r;q))\, dr ,
\ee
where
\be
\rho_{\ga}(r;q):= \sum_{q_i\in q} J_{\ga}(r, q_i) 
\ee
is the local particle density at $r\in\mathbb{R}^d$.
The local density is defined through Kac potentials,
$J_{\gamma}(r,r') = \gamma^d J(\gamma r,\gamma r')$, where $J(s, t)$ is a symmetric, translation invariant ($J(s, t) = J(, t-s)$) smooth 
function which 
vanishes for $|t-s| \geq 1$.
Thus, the range of the interaction has order $\gamma^{-1}$ (for both repulsive and attractive potentials) 
and the ``Kac scaling parameter''  $\gamma$ is assumed to be small. 
This choice of the potentials makes the LMP model a perturbation of the mean-field, in the sense that when taking the thermodynamic limit
followed by the limit $\ga \to 0$ the free energy is equivalent to the free energy in the mean-field description \eqref{mfield}.

Note that the LMP 
 interaction is the sum of a repulsive four body potential
and an attractive two body potential,
which can be written in the following way
\be
H^{\text{LMP}}_{\la,\ga}(q)= - \la |q| - \frac{1}{2!} \sum_{i\neq j} J^{(2)}_{\ga}(q_i,q_j) + \frac{1}{4!} \sum_{i_1\neq ... \neq i_4}
J^{(4)}_{\ga}(q_{i_1},...,q_{i_4}), 
\ee
where  $|\cdot|$ denotes the cardinality of a set and
\begin{align}
&J^{(2)}_{\ga}(q_i,q_j)= \int J_{\ga}(r,q_i) J_{\ga}(r,q_j) \,dr \\ \notag
&J^{(4)}_{\ga}(q_{i_1},...,q_{i_4})= \int J_{\ga}(r,q_{i_1}) \cdots J_{\ga}(r,q_{i_4})\, dr.
 \end{align}

To this model we add an extra hard-core interaction described by 
a potential $V^{R}: \mathbb{R}^d \to \mathbb{R}$
such that
\be\label{hc}
V^{R}(q_i,q_j)= 
\begin{cases}
+\infty \qquad \text{if } |q_i-q_j| \leq R\\
0 \qquad \text{if } |q_i-q_j| > R
\end{cases}
\ee
where $|q_i-q_j|$ denotes the euclidean distance between the two particles in $q_i$ and $q_j$. $R$ is the radius of the hard spheres and their volume is
$\eps=V_d(R)$, i.e., the volume of the d-dimensional sphere of radius R.
Note also that the hard-core potential depends on $q_i, q_j$ only through their distance.

Hence, the Hamiltonian of the model (LMP-hc) we consider is the following
\be
H_{\ga,R, \lambda}(q)=\int e_{\la}(J_{\ga} \ast q(r))\, dr + H^{\text{hc}}_R(q),
\ee
where
\be
 H^{\text{hc}}_R(q):= \sum_{i<j} V^{R}(q_i,q_j).
\ee
Given two configurations $q$ and $\bar q$, we will use the following two notations
to represent the energy of the particle configuration $q$ in the field generated by $\bar q$
and the interaction energy between the particle configuration $q$ and $\bar q$ 
\be\label{hamint}
H_{\ga,R,\la}(q|\bar q)= H_{\ga,R,\la}(q + \bar q)-H_{\ga,R,\la}( \bar q)
\ee
\be\label{int}
U_{\ga,R,\la}(q,\bar q) = H_{\ga,R,\la}(q + \bar q) - H_{\ga,R,\la}(  q) - H_{\ga,R,\la}( \bar q).
\ee
respectively, both for configurations with finitely many particles.

The grand-canonical Gibbs measure in the bounded measurable region $\La$ in $\mathbb{R}^d$ and boundary conditions
$\bar q \in \QQ^{\La^c}$ is the probability measure on $\QQ^{\La}$ defined by
\be\label{gibbsm1}
\mu^{\La}_{\ga,\beta,R,\la, \bar q}(dq)= Z_{\ga, \beta, R, \la, \bar q}^{-1} (\La) e^{
-\beta H_{\ga, R,\la}(q|\bar q)} \nu^{\La}(dq),
\ee
where $\beta$ is the inverse temperature, $ \nu^{\La}(dq)$ is the Poisson point process of intensity $1$ and 
$Z_{\ga, \beta, R, \la, \bar q}(\La)$ is the 
grand canonical partition
function (defined as the normalization factor for $\mu^{\La}_{\ga,\beta,R,\la, \bar q}(dq)$ to be a probability).


\subsection{Mean-field model}\label{sec:mf}
The model introduced above is a perturbation of a mean-field model, which is defined as follows. We
consider the space of configurations with hard-core constraint
\be\label{hc1}
  \mathcal X_{n,\La}^R=
\{ (q_1,...q_n) \in \La^n : \min_{i \neq j} |q_i-q_j| >R \}.
\ee
Given a configuration $q\equiv(q_1,...,q_n)$ in  $\mathcal X_{n,\La}^R$,
the mean-field Hamiltonian is
\be\label{en}
H_{\La,R, \la}^{\text{mf}}(q)= |\La| e_{\la}(\rho)
\ee
where $\rho=n/|\La|$ and $e_{\la}(\cdot)$ is given in \eqref{mfield}. 
The mean-field canonical partition function is
\begin{eqnarray}
Z^{\text{mf}}_{n,\La,R}= \frac{1}{n!} \int_{\mathcal X_{n,\La}^R} e^{-\beta H_{\La,R, 0}^{\text{mf}}(q)}dq_1 \cdots dq_n\\
=\exp\Big\{-\beta\Big( - \frac{n^2}{2|\La|} + \frac{n^4}{4!|\La|^3}\Big)\Big\}\frac{1}{n!} \int_{\mathcal X_{n,\La}^R}dq_1 \cdots dq_n  .
\end{eqnarray}
 The existence of its thermodynamic limit  
follows from general arguments and the canonical mean-field free energy is
\be\label{mfphi}
\phi_{\beta,R} (\rho)= \lim_{|\La|, n \to \infty: \frac{n}{|\La|} \to \infty}
-\frac{1}{ |\La|\beta} \log Z^{\text{mf}}_{n,\La,R} =
e_0(\rho) + f_{\beta,R}^{\text{hc}}(\rho)
\ee
where 
\be\label{free}
 f^{\text{hc}}_{\beta,R}(\rho):=  \lim_{|\La|, n \to \infty: \frac{n}{|\La|} \to \infty}
-\frac{1}{|\La|\beta} \log Z^{\text{hc}}_{n,\La,R}, \qquad 
  Z^{\text{hc}}_{n,\La,R}:= \frac{1}{n!} \int_{\mathcal X_{n,\La}^R} dq_1 \cdots dq_n
 \ee
 is a convex function of $\rho$. 

The mean-field model shows a phase transition for $\beta$ large enough, which is reflected in 
a loss of convexity of $\phi_{\beta,R} (\rho)$. The critical points of $\phi_{\beta,R,\la} (\rho)=\phi_{\beta,R} (\rho)-\la\rho$, as
a function of $\rho$, are the solutions of the mean-field equation
\be
\frac{d}{d\rho} \Big\{ e_{\la}(\rho) +   f^{\text{hc}}_{\beta,R}(\rho) \Big\}=0
\ee
and have the form
\be
\rho= \exp\Big\{ - \beta e_{\la}'(\rho) -  \psi_{\beta,R} ' (\rho)\Big\}:=
K_{\beta, \la, R}(\rho),
\ee
where $\psi_{\beta, R}(\rho)$ is the free energy minus the entropy of the free system, i.e.,
\be
f^{\text{hc}}_{\beta,R}(\rho)-\frac{1}{\beta}\rho(\log\rho-1).
\ee
We have the following properties
\begin{itemize}
\item There is a critical inverse temperature $\beta_{c,R}$,
 such that $ \phi_{\beta,R} (\rho)$ is convex for $\beta \leq \beta_{c,R}$, while for $\beta > \beta_{c,R}$
it has two inflection points $0 < s_{-}(\beta) < s_{+}(\beta)$, being concave for $\rho\in (s_-(\beta), s_+(\beta))$
and convex for $\rho\notin (s_-(\beta), s_+(\beta))$.

\item For any $\beta > \beta_{c,R}$, there is $\la(\beta,R)$ so that $\phi_{\beta, \la(\beta,R), R} (\cdot)$ has two
{global} minimizers, $\rho_{\beta, R, -}<\rho_{\beta, R, +}$ (and a local maximum at $\rho_{\beta, R,0}$). For 
$\la\neq\la(\beta,R)$ and for $\beta\leq \beta_{c,R}$ the minimizer is unique.

\item For any $\beta > \beta_{c,R}$ there is an interval $(\la_-(\beta,R), \la_+(\beta,R))$ 
containing $\la(\beta, R)$ and for any $\la$ in the interval $\phi_{\beta,\la,R}(\cdot)$
it has two local minima $\rho_{\beta,\la,R,\pm}$ which are differentiable functions
of $\la$ and $\frac{d}{d\la}(\phi_{\beta,\la,R}(\rho_{\beta,\la,R,+}) -
\phi_{\beta,\la,R}(\rho_{\beta,\la,R,-}))= \rho_{\beta,\la,R,-} - \rho_{\beta,\la,R,+ }<0$. 
For all $\beta > \beta_{c,R}$,
\be\label{mf2}
\frac{d}{d\rho} K_{\beta, \la(\beta,R), R}(\rho) \Big|_{\rho=\rho_{\beta,R,\pm}}
\equiv K'_{\beta, \la(\beta,R), R}(\rho_{\beta,R,\pm})<1,
\ee
the condition \eqref{mf2} being equivalent to $\phi_{\beta,\la(\beta),R} '' (\rho_{\beta,R,\pm})>0$.
Moreover, there exists $\beta_{0,R}>\beta_{c,R}$ such that
\be\label{mf22}
K'_{\beta, \la(\beta,R), R}(\rho_{\beta,R,\pm})>-1, \qquad \text{for all } \beta \in (\beta_{c,R}, \beta_{0,R}).
\ee 

\item We have an expansion for $\beta_{c,R}$ in powers of $\eps=V_d(R)$
\[
\beta_{c,R}=\beta^{\text{LMP}}_c-  \eps \,(\beta^{\text{LMP}}_c)^{2/3} + O(\eps^2),
\]
$ \beta^{\text{LMP}}_c=\frac 3 2^{\frac 3 2}$ being the critical inverse temperature for the LMP mean-field model.
Note that while $\beta_{c,R}$ has the meaning of critical inverse temperature, $\beta_{0,R}$ has no physical meaning, 
but it is introduced for technical reasons. In fact $\beta_{0,R}$ is necessary for \eqref{mf22} to be true 
and depends on the choice of the mean-field Hamiltonian \eqref{en}.

\end{itemize}

\section{Contour model}\label{sec:con}

To prove the phase transition in the LMP-hc model we study perturbations 
of the homogeneous states with densities $\rho_{\beta,R,\pm}$ which appear in the limit $\ga\to 0$.
We follow an argument \`a la  Peierls, 
which relies (as for the Ising model) on the possibility to rewrite
 the partition function of the model as the partition function of an ``abstract contour model''.
 To implement this strategy we need to introduce several scaling parameters and phase indicators. Namely, we introduce
 two scales $\ell_{\pm}= \ga^{-(1 \pm \alpha)}$ and an accuracy parameter $\zeta= \ga^a$, with $ 1\gg \alpha\gg a>0$.
We define $\mathcal D^{(\ell)}$ a partition of $\mathbb R^d$ into cubes of side $\ell$ and we denote $C_r^{(\ell)}$ the
cube of $\mathcal D^{(\ell)}$ which contains $r$.

The first phase indicator is defined as
\[
 \eta^{(\zeta,\ell_-)}(q;r)= \begin{cases}
\pm 1 \qquad \text{if }  \Big|   \rho^{(\ell_-)}(q;r)  - \rho_{\beta,R,\pm}\Big|
\le \zeta\\
0 \qquad \text{otherwise } 
\end{cases}
\]
where $ \rho^{(\ell)}(q;r) =|C_r^{(\ell)}\cap q|\ell^{-d}$ is the empirical densiy in a cube of side $\ell$ 
containing $r$ given a configuration $q$.

Thus $\eta^{(\zeta,\ell_-)}(q;r)$ indicates the phase (or its absence) on the small scale $\ell_-$.
Because of statistical fluctuations, we must allow for deviations
from the ideal plus configurations $\eta^{(\zeta,\ell_-)}(q;r)=1$. 
We thus need to define which regions are still in the
plus phase and which are those destroyed by the fluctuations.
The fact that $\eta^{(\zeta,\ell_-)}(q;r)=1$ does not
qualify $r$ being in the $+$ phase, implies that we need
a stronger condition which is defined in terms of two more phase indicators
which describe the local phase of the system in increasing degree of accuracy. We have
\[
\theta^{(\zeta,\ell_-,\ell_+)}(q;r)= \begin{cases}
\pm 1 \qquad \text{if }  \eta^{(\zeta,\ell_-)}(q;r')=\pm 1 
\quad \forall r' \in C^{(\ell_+)}_r\\
0 \qquad \text{otherwise } 
\end{cases}
\]
\[
\Theta^{(\zeta,\ell_-,\ell_+)}(q;r)= \begin{cases}
\pm 1 \qquad \text{if }  \eta^{(\zeta,\ell_-)}(q;r')=\pm 1 
\quad\forall r' \in C^{(\ell_+)}_r\cup
\delta^{\ell_+}_{\text{out}}[C^{(\ell_+)}_r]\\
0 \qquad \text{otherwise } 
\end{cases}
\]
where $\delta^\ell_{\rm out}[\La]$ of a
$\mathcal D^{(\ell)}$-measurable region $\La$  is the union
of all the cubes $C\in \mathcal D^{(\ell)}$ next to $\La$. 
For simplicity, from now on we drop the superscript from the notation of $\eta^{(\zeta,\ell_-)}, \theta^{(\zeta,\ell_-,\ell_+)},
\Theta^{(\zeta,\ell_-,\ell_+)}$.

With these definitions, given a configuration $q$, the ``plus phase'' is 
the region $\{r : \Theta(q;r)=1\}$ while the ``minus phase''
is the region $\{r : \Theta(q;r)=-1\}$.
We call $q^{\pm} $ a $\pm$ boundary conditions relative to a region $\La$, if it belongs to the ensemble
$\eta(q;r)=\pm1$ for $r$ on the frame of width $2\ga^{-1}$ around $\La$.

Two sets are connected if their 
closures have non empty
intersection; hence, two cubes with a common vertex are connected. 
In this way, the plus and the minus regions are separated by zero-phase regions 
$\{r : \Theta(q;r)=0\}$.

  \begin{definition}
A contour   is a  pair $\Ga= \big(\text{sp}(\Ga),\eta_\Ga\big)$,
where ${\rm sp}(\Ga)$
is a maximal connected component of the ``incorrect set'' $\{r\in \mathbb R^d:
\Theta(q;r)=0\}$ and $\eta_\Ga$ is the
restriction to $\text{sp}(\Ga)$
of $\eta(q;\cdot)$.

\end{definition}
The exterior, ${\rm ext}(\Ga)$, of  $\Ga$
is the unbounded, maximal connected component of ${\rm
sp}(\Ga)^c$. The interior  is the set
${\rm int}(\Ga)={\rm sp}(\Ga)^c\setminus {\rm ext}(\Ga)$;
we denote by ${\rm int}_i(\Ga)$ the maximal connected
components of ${\rm int}(\Ga)$.
 Let
$c(\Ga)  = \text{sp}(\Ga) \cup \text{int}(\Ga)$
and note that
${\rm int}_i(\Ga)$  and $c(\Ga)$ are both simply connected.
The outer boundaries of $\Ga$ are the sets
    \begin{equation}
      \label{p4.3.4.1}
A(\Ga):=\delta_{\text{\rm
out}}^{\ell_+}[\text{sp}(\Ga)]\cap {\rm int}(\Ga),\;
A_{\rm ext}(\Ga):=\delta_{{\rm out}}^{\ell_+}[c(\Ga)].
     \end{equation}
We will also call $A_i(\Ga)=A(\Ga)\cap {\rm int}_i(\Ga)$.

\begin{definition}
$\Ga$ is a plus/minus, contour if
$\Theta(q;r)=\pm 1$ on  $A_{\rm ext}(\Ga)$. 
\end{definition}

We add a superscript $\pm$ to $A_i(\Ga)$ to indicate the sign of $\Theta$
and we write ${\rm int}^{\pm}_i(\Ga)$ if ${\rm int}_i(\Ga)$ contains $A^{\pm}_i(\Ga)$.  
Note that $\Theta$ is constant on $A_{\rm ext}(\Ga)$ and $A_i(\Ga)$ and its value
is determined by $\eta$.

  \begin{definition}
Given a plus contour $\Ga$ and a plus boundary condition
$q^+$ for $c(\Ga)$, we define the weight
$W^{+}_{\ga,R,\la}(\Ga;\bar q)$ of
$\Ga$ as equal to
    \begin{equation}
    \label{9.1a.4.1}
 \frac{
\mu^{c(\Ga)}_{\ga,\beta,R,\la,q^+}\Big(\eta(q_{c(\Ga)};r)=
\eta_\Ga(r), r \in {\rm sp}(\Ga);\;
\Theta(q_{c(\Ga)};r)=\pm 1, r\in A^{\pm}(\Ga) \Big)}
{ \mu^{c(\Ga)}_{\ga,\beta,R,\la,q^+}
\Big(\eta(q_{c(\Ga)};r)=1, r \in {\rm
sp}(\Ga);\; \Theta(q_{c(\Ga)};r)=  1, r\in A^{\pm}(\Ga)
\Big)}
     \end{equation}
     where the measure $\mu^{c(\Ga)}_{\ga,\beta,R,\la,q^+}$ 
     has been defined in \eqref{gibbsm1}. Analogously, we can
     define the weight of a minus contour.
\end{definition}
Thus, the numerator is the probability of the contour $\Ga$ conditioned 
to the outside of $\rm sp(\Ga)$ while the denominator is the probability 
that the contour $\Ga$ is absent and replaced by the plus configurations
(with the same conditioning to the outside).

The weight  $W^{-}_{\ga,R,\la}(\Ga;q^{-})$ of a minus contour  $\Ga$ is defined analogously.
The weight  $W^{\pm}_{\ga,R,\la}(\Ga; q^{\pm})$ depends only
on $q^{D}$, i.e., the restriction of $q^{\pm}$ to $D\equiv\{r\in c(\Ga)^c: {\rm
dist}(r,c(\Ga))\le 2\ga^{-1}\}$.

\begin{definition}
 The plus  diluted Gibbs measure  in
a bounded  $\mathcal
D^{(\ell_{+})}$-mea\-su\-rable region $\La$ with plus
boundary conditions $\bar q$ is
   \begin{equation}
    \label{9.1a.2.1}
\mu^{\La,+}_{\ga,\beta,R,\la,\bar q}(dq_{\La}):=
\frac{1}{Z^{+}_{\ga,\beta,R,\la,\bar q}(\La) }e^{-\beta
H_{\ga,R,\la}(q_\La|\bar q_{\La^c})} \text{\bf
1}_{\Theta((q_\La+q^{+}_{\La^c});r)=1\;\text{$r\in
\delta_{\text{\rm out}}^{\ell_{+}} [\La^c]$}}
\nu^\La(dq_\La)
     \end{equation}
where $q^+\in\QQ^+=\{q: \eta(q;r)=1, r\in\mathbb R^d\}$ and $Z^{+}_{\ga,\beta,R,\la,\bar q}(\La) $
is the normalization, also called the plus diluted partition function. A similar definition holds for 
the minus  diluted Gibbs measure.
\end{definition}

We end this section by writing the
ratio  \eqref{9.1a.4.1}
of probabilities in the definition of the weight of a
contour as a ratio of two partition
functions. By writing explicitely the contributions coming from the support of a contour and those coming 
from the interior, we have for  
a plus contour $\Ga$ 
       \begin{eqnarray}
&&  W^{+}_{\ga,R,\la}(\Ga;q^{+}) = \frac{\mathcal
N^+_{\ga,R,\la}(\Ga,q^+)}{\mathcal D^+_{\ga,R,\la}(\Ga,q
 ^+)}
    \label{9.1c.1.1.1}
     \end{eqnarray}
where:
       \begin{eqnarray}
 \mathcal N^+_{\ga,R,\la}(\Ga,q^{+}) &=&  \int_{q_{{\rm sp}(\Ga)}:
 \eta(q_{{\rm sp}(\Ga)} ;r) = \eta_\Ga (r), r \in
{\rm sp}(\Ga)} \;\;e^{-\beta
H_{\ga,R,\la, {\rm sp}(\Ga)}(q_{{\rm sp}(\Ga)
}|q^{+}_{A_{\rm ext}})}
 \nonumber
\\
&\times&
{Z}^{-}_{\ga,\beta,R,\la,q_{{\rm sp}(\Ga)}}(
{\rm int}^{-}(\Ga)) \,{Z}^{+}_{\ga,\beta,R,\la,q_{{\rm sp}(\Ga)}}(
{\rm int}^{+}(\Ga))
    \label{9.1d.1.3}
     \end{eqnarray}
       \begin{eqnarray}
 {\mathcal
D}^{+}_{\ga,R,\la}(\Ga,q^{+}) &=&  \int_{q_{{\rm sp}(\Ga)}:
 \eta(q_{{\rm sp}(\Ga)} ;r) = 1, r \in
{\rm sp}(\Ga)} \;\;e^{-\beta
H_{\ga,R,\la, {\rm sp}(\Ga)}(q_{{\rm sp}(\Ga)
}|q^{+}_{A_{\rm ext}})}
 \nonumber
     \\
&\times&
{Z}^{+}_{\ga,\beta,R,\la,q_{{\rm sp}(\Ga)}}(
{\rm int}^{-}(\Ga))\, {Z}^{+}_{\ga,\beta,R,\la,q_{{\rm sp}(\Ga)}}(
{\rm int}^{+}(\Ga)).
     \label{9.1d.1.4}
     \end{eqnarray}

\section{The main results}\label{sec:main}
Our main theorem states that the system undergoes a first-order phase transition. This means that
for $\beta$ large enough the Gibbs state at the thermodynamic limit, i.e. $\La\to\mathbb R^d$, is not
unique. It is possible to fix plus/minus boundary conditions such that, if $R$ and $\ga$ are small and for some
values of $\beta,\la$, uniformly in $\La$, the typical configurations of the
corresponding diluted Gibbs measures are close to the plus/minus phase. 
This is quantified in the following theorem.

\begin{theorem}[Liquid-vapor phase transition]\label{mainthm}

Consider the LMP-hc model in dimensions $d\geq 2$.
For such a model there are $R_0$, $\beta_{c,R}, \beta_{0,R}$ 
and for any $0<R\leq R_0$ and $\beta \in (\beta_{c,R},\beta_{0,R})$ there is $\ga_{\beta,R}>0$ 
so that for any $\ga \leq \ga_{\beta,R}$ there is $\la_{\beta,\ga,R}$ such that:
\vspace{0.5cm}

\noindent
 There are two distinct infinite-volume measures $\mu_{\beta,\ga,R}^{\pm}$ with chemical potential 
$\la_{\beta,\ga,R}$ and inverse temperature $\beta$ and two different densities:
 $0 < \rho_{\beta,\ga,R,-}<\rho_{\beta,\ga,R,+}$.
  \end{theorem}
  
 In the theorem, $\mu_{\beta,\ga,R}^{\pm}$ are the infinite-volume limits of \eqref{9.1a.2.1}, 
 while $\beta_{c,R},\beta_{0,R}$ are the two inverse temperatures introduced in Section \ref{sec:mf}. 
 
We prove the existence of two distinct states, which are interpreted as the two pure phases of the system: 
$\mu_{\beta,\ga,R}^{+}$ describes the liquid phase with density $\rho_{\beta,\ga,R,+}$ 
while $\mu_{\beta,\ga,R}^{-}$ describes the vapor phase, with the smaller density $\rho_{\beta,\ga,R,-}$.
Furthermore we have
\[
\lim_{\ga\to 0} \rho_{\beta,\ga,R,\pm} = \rho_{\beta,R,\pm}, \quad \rho_{\beta,R,-}< \rho_{\beta,R,+},
\qquad
\lim_{\ga\to 0} \la_{\beta,\ga,R} =\la(\beta,R)
\]
which are the densities and the chemical potential for which there is a phase transition
in the mean-field model (see again Section \ref{sec:mf}). 

The main technical point in the proof of Theorem \ref{mainthm} is to
  prove that contours are improbable. In particular, they satisfy Peierls estimates 
  which proves that the probability of a contour decays exponentially with its volume.  

\medskip

\begin{theorem}\label{thm:P}

There exists $R_0$ such that for any $R\leq R_0$ and any $\beta\in (\beta_{c,R},\beta_{0,R})$  there exist $c>0$, 
$\ga_{\beta,R}>0$,  so that for any $\ga\le
\ga_{\beta,R}$, $\pm$ contour $\Ga$ and any
$\pm$ boundary condition $q^{\pm}$ relative to $c(\Ga)$,
     \begin{equation}
    \label{9.1a.5.1}
W^{\pm}_{\ga,R,\la}(\Ga;q^{\pm}) \le \exp\Big\{ -\beta
c \, (\zeta^2 \ell_{-}^d)\,  N_{\Ga }
 \Big\}
     \end{equation}
     where $\la=\la_{\beta,\ga,R}$ and
     \be
N_\Ga = \frac{|{\rm sp}(\Ga)|}{\ell_{+}^d}
\ee
is the number of cubes of the partition $\mathcal D^{(\ell_+)}$ contained in $\rm sp(\Ga)$.
   \end{theorem}
   
   As a corollary of Theorem \ref{thm:P} we have

   \begin{corollary}
   \label{thm:cor}

There exists $R_0$ such that for any $R\leq R_0$, any $\beta\in (\beta_c,\beta_0)$
and
letting $c$, $\ga_{\beta,R}$,  $\ga$ and $\la_{\beta,\ga,R}$ as in
Theorem \ref{thm:P}, we have that for any bounded, simply
connected, $\mathcal D^{(\ell_{+})}$ measurable region
$\La$, any $\pm$ boundary condition $q^{\pm}$ and
any $r\in \La$, the following holds
    \begin{equation}
    \label{p8.1.4.3a.00}
\mu^{\La,\pm}_{\ga,\beta,R,\la_{\beta,\ga,R},q^{\pm}}( \{\Theta(q;r)=\pm1\}) 
\ge \; 1\;-\;  \exp\Big\{ -\beta \frac{c}{2} \,
(\zeta^2 \ell_{-}^d)\Big\}.
     \end{equation}

        \end{corollary}

Theorem \ref{thm:cor} implies that  for any $R\leq R_0$ 
and $\ga$ small enough (chosen according to $R$) 
the difference between the diluted Gibbs measures
$\mu^{\La,+}_{\ga,\beta,R,\la_{\beta,\ga,R},q^{+}}(dq)$ and
$\mu^{\La,-}_{\ga,\beta,R,\la_{\beta,\ga,R},q^{-}}(dq)$ survives in the
thermodynamic limit $\La\nearrow \mathbb R^d$ and a phase
transition occurs.  

The main difficulty in proving \eqref{9.1a.5.1} is that both numerator and 
denominator in \eqref{9.1a.4.1} are defined in terms of expressions which involve not only
the support of $\Ga$ but also its whole interior. They are therefore ``bulk quantities'' while the desired 
bound involves only the volume of the support of $\Ga$, which for some contours, at least, is a ``surface quantity''.
The main issue here is to find  cancellations of the bulk terms between the numerator and the denominator. 
This is easy when special symmetries allow to relate the $+$ and $-$ 
ensembles, as in the ferromagnetic Ising model. Such simplifications are not present here and this is one of the issues which makes continuum models difficult to study. 
We overcome this difficulty using the Pirogov-Sinai theory \cite{PS} which covers cases where 
this symmetry is broken.

A central point of the Pirogov-Sinai theory is a change of measure. The idea
is to introduce a new Gibbs measure (simpler than the original one),
but which gives the same properties. 
The diluted partition function
in a region $\La$ can be written as a partition function in  $\mathcal Q_+^\La=\{ q\in \mathcal Q_\La:
\eta(q,r)=1, r\in \La\}$. Namely, for any bounded $\mathcal D^{(\ell_{+})}$-measurable
region $\La$ and any plus b.c.\  $q^+$, we have that
      \begin{equation}
         \label{a8z.3.1.4}
Z^{+}_{\ga,\beta,R,\la,q^+}(\La) = \sum_{\und\Ga\in \mathcal
B^{+}_\La} \int_{q_\La\in \mathcal
Q_+^\La}W^+_{\ga,R,\la}(\und\Ga,q_\La)\; e^{-\beta
H_{\ga,R,\la}(q_\La |q^+_{\La^c})},
    \end{equation}
   where $q^+_{\La^c}$ is made of all 
particles of $q^+$ which are in $\La^c$. $\mathcal B_{\La}^+$ is the space of all finite
subsets  of collection of plus contours made of  elements which are
mutually disconnected and with spatial support not connected to $\La^c$. Furthermore  
if  $\und \Ga=(\Ga_1,..,\Ga_n)$, we use the notation
          \begin{equation}
         \label{9.1d.1.1.1}
W^{\pm}_{\ga,R,\la}(\und\Ga,q)= \prod_{i=1}^n 
W^{\pm}_{\ga,R,\la}(\Ga_i,q).
    \end{equation}
A similar expression holds for the diluted minus  partition function.

In order to prove Peierls bounds, we follow the version of the Pirogov-Sinai theory proposed
by Zahradnik \cite{Z}. In this picture large contours are less likely to be observed and
this is implemented by fixing a constraint which literally forbids contours larger than some
given value. We introduce therefore a 
new class of systems, where the contour weights are modified,
their values depending on some ``cutoff'' parameter. In the stable phase the 
cutoff (if properly chosen) is not reached and the state is not modified by
this procedure. 

%


Therefore we
choose  ${\hat W}^{\pm}_{\ga,R,\la}(\Ga;q^{\pm})$, 
positive numbers which depend only on the restriction of
$q^{\pm}$ to $\{r\in
c(\Ga)^c:{\rm dist}(r,c(\Ga))\le 2 \ga^{-1}\}$ and
such that
for any $\pm$ contour $\Ga$ and any $q^{\pm}$,
      \begin{eqnarray}
&&  {\hat W}^{\pm}_{\ga,R,\la}(\Ga;q^{\pm}) = \min\Big\{
\frac{\hat{\mathcal N}^{\pm}_{\ga,R,\la}(\Ga,q)}{\hat{
 \mathcal D}^{\pm}_{\ga,R,\la}(\Ga,q)},  e^{ -\beta \frac{c}{100} \,
(\zeta^2 \ell_{-}^d)\,  N_{\Ga } }\Big\}.
    \label{9.1d.1.5}
     \end{eqnarray}
Here, $\hat{\mathcal N}^{\pm}_{\ga,R,\la}(\Ga,q)$ and $\hat{\mathcal
D}^{\pm}_{\ga,R,\la}(\Ga,q)$ are as in
\eqref{9.1d.1.3}-\eqref{9.1d.1.4} but depend uniquely on the weights ${\hat 
W}^{\pm}_{\ga,R,\la}(\cdot;\cdot)$.
With this new choice of contours weights, if we prove Peierls bounds, i.e.,
\eqref{9.1a.5.1} on definition \eqref{9.1d.1.5}, we have Peierls bounds also
on the ``true'' weights defined in \eqref{9.1c.1.1.1}. 
We write $\hat Z^{+}_{\ga,\beta,R,\la,q^+}(\La)$ to denote the new diluted partition
function. For more details one can see in \cite{Pre}.

\section{Outline of the proof}
In this section we want to give a sketch of the proof of \eqref{9.1a.5.1}
for the case of the cutoff contours as defined above. For the complete 
proof of the argument see \cite{Pre}. 

The first step is to prove that 
it is possible to separate in \eqref{9.1d.1.3}-\eqref{9.1d.1.4} 
the estimate in ${\rm int}(\Ga)$
from the one in ${\rm sp}(\Ga)$ with ``negligible error''. 
Then one needs to bound  a constrained partition
function in ${\rm sp}(\Ga)$,  which yields the gain factor $e^{-\beta
(c\zeta^2 - c'\ga^{1/2-2\alpha d})\ell_{2}^d N_\Ga}$. 
Hence, we prove that there are $c, c'>0$ so that given $\ga$ small enough, for $R<R_0$,

        \be
 \frac{{\hat{\mathcal
N}}^+_{\ga,R,\la_{\beta,R,\ga}}(\Ga, q^+)}{{\hat{
\mathcal D}}^{+}_{\ga,R,\la_{\beta,R,\ga}}(\Ga, q^+)} \le 
e^{-\beta
(c\zeta^2 - c'\ga^{1/2-2\alpha d})\ell_{2}^d N_\Ga}
\frac{e^{ \beta I^-_{\ga,\la(\beta,R)}({\rm int}^-(\Ga))}
{\hat Z}^{-}_{\ga,R,\la_{\beta,\ga,R},
\chi^-}({\rm int}^-(\Ga)) } {e^{ \beta
I^+_{\ga,\la(\beta,R)}({\rm int}^-(\Ga))}{\hat Z}^{+}_{\ga,R,\la_{\beta,\ga,R},
\chi^+}({\rm int}^-(\Ga)) }
    \label{cazzo}
     \ee
where we use the shorthand
notation
      \begin{eqnarray}
&&  \chi^{\pm}_{\Delta}(r)=\rho_{\beta,\pm}  \text{\bf
1}_{r\in \Delta}, \quad \chi^{\pm}=\chi^{\pm}_{\mathbb R^d}
    \label{9b.4ccc}
     \end{eqnarray}
     and where $I^{\pm}_{\ga,\la(\beta,R)}(\La)$ is a surface term
       \begin{eqnarray}
&&   I^{\pm}_{\ga,\la(\beta,R)}(\La)= \int_{\La^c}
\{e_{\la(\beta,R)}(\rho_{\beta,R,\pm})- e_{\la(\beta,R)}(J_\ga *
\rho_{\beta,R,\pm}\text{\bf 1}_{\La^c})\}\\
&&\hskip2cm
-\int_{\La}e_{\la(\beta,R)}(J_\ga * \rho_{\beta,R,\pm}\text{\bf
1}_{\La^c}).
    \label{9.1c.3.2.1}
     \end{eqnarray}

The main tool used in this part of the proof is a coarse-graining argument and an analysis
\`a la Lebowitz and Penrose, \cite{LP}. The
error in doing a coarse-graining is bounded by $e^{\beta c\ga^{1/2}|{\rm sp}(\Ga)|}=
e^{\beta c\ga^{1/2-2\alpha d}\ell_{2}^d N_\Ga}$, which is
the ``negligible factor'' mentioned above, as it is a small fraction  of the gain term in the Peierls bounds.
Thus, in this step we have a reduction, after coarse-graining, to variational problems
with the LMP free energy functional. They involve two different regions, one is at the 
boundary between ${\rm int}(\Ga)$ and ${\rm sp}(\Ga)$, the other is in the bulk of 
the spatial support. 
In the former we exploit
the definition of contours which implies that the
boundary of int$^{\pm}(\Ga)$ is in the middle of a ``large
region'' (of size $\ell_{+}$) where $\eta(\cdot;\cdot)$
is identically equal to $\pm 1$, respectively. By the
strong stability properties of the LMP free energy
functional, the minimizers are then proved to converge
exponentially to $\rho_{\beta,R,\pm}$ with the distance from
the boundaries. Here we use the assumption that $\beta\in
(\beta_{c,R},\beta_{0,R})$, i.e., where the mean-field operator
$K_{\beta, \la(\beta,R), R}$ is a contraction, see \eqref{mf2}-\eqref{mf22}. 
 We then conclude that with a
negligible error we have   ``thick corridors'' where the
minimizers are equal to  $\rho_{\beta,R,\pm}$  thus
separating the regions outside and inside the corridors.

After this step we have plus/minus partition
functions in  int$^{\pm}(\Ga)$ with boundary conditions
$\rho_{\beta,R,\pm}$ and still a variational problem  in the
region ${\rm sp}(\Ga)$ with the constraint that profiles should be
compatible with the presence of
the contour $\Ga$. The analysis of such a minimization
problem leads to the gain factor in the Peierls bounds.

To complete the proof for Peierls bounds we then need to prove the 
following theorem.

\begin{theorem}\label{thm:surcor}
There exists $R_0$ such that for any $R\leq R_0$ and any $\beta\in (\beta_{c,R},\beta_{0,R})$  there are $c>0$, 
$\ga_{\beta,R}>0$ and  $\la_{\beta,\ga,R}$, such that for all $\ga\le
\ga_\beta$, $|\la(\beta,R)-\la_{\beta,R,\ga}|\le c\ga^{1/2}$, 
 and any bounded $\mathcal
D^{(\ell_{+})}$-measurable region $\La$, the following bound holds
       \begin{eqnarray}
&& \frac{ e^{\beta I^-_{\ga,\la(\beta,R)}(\La) }
{\hat Z}^-_{\ga,R,\la_{\beta,\ga,R}, \chi^-_{\La^c}} (\La)}{ e^{\beta
I^+_{\ga,\la(\beta,R)}(\La)}{\hat Z}^{+}_{\ga,R,\la_{\beta,\ga,R},\chi^+_{\La^c}}(\La) } 
\;\le \; e^{c\ga^{1/2} |\delta_{{\rm
out}}^{\ell_{+}}[\La]|}.
    \label{9.1c.4.1.11n}
     \end{eqnarray}
\end{theorem}

The idea in the proof of \eqref{9.1c.4.1.11n} is that the leading term in the 
partition function is 
\be\label{rozza}
 {\hat Z}^{\pm}_{\ga,\beta,R,\la,q_{{\rm sp}(\Ga)}}(
{\rm int}^{\pm}(\Ga))\approx
e^{\beta P^{\pm}_{\ga, R, \la} |{\rm int}^{\pm}(\Ga)|},
\ee
where $P^{\pm}_{\ga, R, \la}$ is the thermodynamic pressure
given, for any van Hove sequence of $\mathcal D^{(\ell_+)}$-measurable regions $\La_n$ 
and any $\pm$ $\La_n$-boundary conditions $q^{\pm}_n$, by the following limit
\be\label{limitpre}
\lim_{n\to\infty} \frac{1}{\beta |\La_n|} \log {\hat Z}^{\pm}_{\ga,R, \la,q^{\pm}_n}(\La_n) = P^{\pm}_{\ga, R, \la}.
\ee
Although \eqref{rozza} is a rough approximation, we need to prove equality of $\pm$ pressures 
in the bulk terms in
$\hat{ W}^{\pm}_{\ga,R,\la}(\Ga;q^{\pm})$ to allow for the numerator and the denominator to cancel.
Again for more details we refer the reader to \cite{Pre}.

We now prove that the next term, i.e., the surface corrections
to the pressure, are small as $e^{c'' \ga^{1/2} \ell_{+}^d N_\Ga}$
at least when the boundary conditions ``are perfect'', i.e., given by $\chi^{\pm}$.
The most difficult step in the proof of Theorem \ref{thm:surcor} are estimates 
involving terms which are localized in the bulk of the interior. These rely on a more delicate 
property of decay of correlations (Theorem \ref{surcorr}), whose proof requires a whole 
new set of ideas.

\begin{theorem}[Exponential decay of correlations]\label{surcorr}
Let $\La$ be a bounded $\mathcal D^{(\ell_+)}$ measurable region.
Let $x_i$ be the centers of the cubes $C^{(\ell_-)} \in \mathcal D^{(\ell_-)}$;
then we define
\be\label{expec}
 f_{x_1,..,x_n} = \int_{\{r_i \in C^{(\ell_-)}_{x_i}, 1\le i\le n\}}
q^{\otimes n}(dr_1..dr_n) J_\ga^{(n)}(r_1,..,r_n)
\ee
where we use the notation
\be\label{mis}
q^{\otimes n}(dr_1... dr_n)= \frac{1}{n!}
\sum_{i_1\neq .. \neq i_n} \delta_{q_{i_1}}(r_1) \,dr_1 \cdots  \delta_{q_{i_n}}(r_n) \,dr_n.
\ee
There are positive constants  $\delta$, $c'$ and $c$   so that for all
$f_{x_1,..,x_n}$
   \begin{equation}
   \label{cruciale}
 \Big| E_{\mu^1} \big(f_{x_1,..,x_n} \big)
-E_{\mu^2}\big(f_{x_1,..,x_n}\big)\Big|
 \le c' e^{-c[\ga^{-\delta} \ell_+^{-1}{\text
dist}(C^{(\ell_-)}_{x_1},\La^c)]}   
   \end{equation}
where $E_{\mu^i}$, $i=1,2$, are the expectations with respect to the 
following two measures: $\mu^1$ is the
 finite-volume Gibbs measure in $\La$ with b.c. $\bar q$ and $\mu^2$ the finite-volume 
 Gibbs measure on a torus $\TT$ much larger than $\La$.

\end{theorem}

We compute the expectations in \eqref{cruciale} in two steps. We first do a coarse-graining by fixing 
the number of particles in the cubes $C^{(\ell_-)}$ and integrate over their positions; then, in the second step,
we sum over the particle numbers. 
By its very nature, the Kac assumption makes the first step simple: in fact, to first order the energy 
is independent of the positions of the particles inside each cube.
Neglecting the higher order terms, the energy drops out of the integrals  (with fixed particle numbers)
which can then be computed explicitly. The result is the phase space volume of the set of configurations with
the given particle numbers: this is an entropy factor which, together with the energy, reconstructs
the mesoscopic energy functional.

By using cluster expansion techniques, we will show here that it is possible to compute
exactly the correction due to the dependence of the energy on the actual positions of the 
particles in each cube. For  the hard-core part of the interaction we can use again a cluster expansion
technique, using the result \cite{PT} obtained for a system with a single canonical constraint
and therefore extending it to the present case of multi-canonical constraints.

Once we are left with an ``effective Hamiltonian'' we still have to sum over the
particle numbers. Since we work in a contour model, the particle
densities are close to the mean-field values $\rho_{\beta,R,\pm}$ so that
the marginal of the Gibbs measure over the coarse-grained model
is Gibbsian and it is a small perturbation of a Hamiltonian given by the mean-field
free energy functional restricted to a neighborhood of the mean-field equilibrium
density.
In such a setup we manage to prove the validity of the Dobrushin 
uniqueness condition, where we take into account the contribution of the hard-core as a cluster expansion sum. 

\subsection{Coarse-graining}
To carry out this plan, we need to
prove that 
$\hat Z^{+}_{\ga,\beta,R,\la,\bar q}(\La)$ can be written as the partition function
of a Hamiltonian which depends on   variables $\rho_x$,  
$x\in X^{(\ell_-)}_\La$, $X^{(\ell_-)}_\La$ the set of  centers of cubes 
$C^{(\ell_-)}$  in $\La$. 
 The new energy of a density configuration
  $\rho=\{\rho_x\}_{x\in X^{(\ell_-)}_{\La}}$ is
defined as
        \begin{equation}
        \label{7.2.2}
h(\rho|\bar q) = -\log  \sum_{\underline\Ga\in \mathcal B^{+}_{\La}}
\int_{\mathcal Q^{\La}_+} \nu^\La(dq)
\text{\bf 1}_{\rho^{(\ell_-)}(q)= \rho } \; e^{-\beta H_{\ga,R,\la}(q|\bar q)} 
\hat W(\underline\Ga|q) 
      \end{equation}
so that
        \begin{equation*}
\hat Z^{+}_{\ga,\beta,R,\la,\bar q}(\La)=   \sum_{\rho}  e^{-
   h(\rho|\bar q) }.
      \end{equation*}
      
      Setting $n_x= \ell_-^d \rho_x$, we  
  multiply and divide, inside the argument of the log
in \eqref{7.2.2}, by
        \begin{equation*}
  \prod_{x\in X_\La} 
\frac{\ell_-^{d n_x}}{ n_x !}.
      \end{equation*}
 We denote by $\{q_{x,i}$,  $i=1,..,n_x$,
$x \in X_\La\}$,  the particles  in $C_x^{(\ell_-)}$.  Thus particles are now
labelled by the pair $(x,i)$, $x$ specifies the cube
$C_x^{(\ell_-)}$ to which the particle ``belongs'', $i$ distinguishes
among the particles in $C_x^{(\ell_-)}$.
The corresponding
 free measure, whose expectation is denoted by
$E^0_{\rho}$, is the product of the probabilities
which give  uniform
distribution to the  positions $q_{x,i}$  
 in their  boxes $C^{(\ell_-)}_x$ divided by $n_x!$
since the particles in each box $C^{(\ell_-)}_x$ are indistinguishable.
Note that when we change from labeling 
of all particles in $\La$ to labeling separately the particles in each
box we have to multiply by $\frac{N!}{\prod_{x\in X^{(\ell_-)}_{\La}}n_x!}$ for
all such possibilities.

 We define a new a priori measure  
 for the particles in a given box $C^{(\ell_-)}_x$, $x \in X_\La$,
as
\be\label{newmeas}
\frac{dq_{x,1} \cdots dq_{x,n_x} e^{-\beta U^{\text{hc}}(q^{(C_x)},\bar q)}}
{\int dq_{x,1} \cdots dq_{x,n_x} e^{-\beta U^{\text{hc}}(q^{(C_x)},\bar q)}}
Z_{x,\bar q}(\rho_x)
\ee
where $q^{(C_x)}$ denotes the configuration of the particles in $C_x^{(\ell_-)}$, 
each integral in the denominator is over $C_x^{(\ell_-)}$ 
with the constraint ${\mathcal Q^{\La}_+}$
and where
\begin{equation}\label{7.2.31}
  Z_{x,\bar q}(\rho_x)=\int_{\mathcal Q^{\La}_+}  \frac{dq_{x,1}}{\ell_-^d}\ldots 
 \frac{dq_{x,n_x}}{\ell_-^d}
 \; e^{-\beta U^{\text{hc}}(q^{(C_x)},\bar q)}
 \end{equation}
is the extra factor coming from the change of measure and  
contributing for each cube
 with
\be\label{boundary}
U^{\text{hc}}(q^{(C_x)},\bar q):=
 \sum_{i=1}^{n_x}\sum_{j=1}^{|\bar q|} V^{R}(q_{x,i}-\bar q_j).
\ee 

The corresponding expectation will be denoted by $E^0_{\rho,\bar q}$.
We then  have
        \begin{align}
        \label{7.2.3}
h( \rho |\bar q) &= - \sum_x \log \frac{\ell_-^{d n_x}}{ n_x !} 
-\sum_{x:\, C_x^{(\ell_-)}\in\partial\La^{\text{int}}} \log Z_{x,\bar q}(\rho_x)\\ \nonumber
&-\log E^0_{\rho,\bar q}\Big(e^{-\beta H_{\ga} (q |\bar q )}e^{-\beta H^{\text{hc}}_R(q)} 
\sum_{\text{sp$(\Ga) \subseteq \La^0$}} 
W(\Ga|q)\Big)
      \end{align}
where $\partial\La^{\text{int}}$ is the set of the $\mathcal D^{(\ell_-)}$
boxes adjacent to $\La^c$ (i.e., the interior boxes of $\La$). 
 Note that the total normalization is a product of the normalizations in each cube
 and that, because of the hard-core interaction, 
 $ Z_{x,\bar q}(\rho_x)$ for a given box $C_x$ gives the following
 contribution:
 \be
  Z_{x,\bar q}(\rho_x)=\Big( \int_{C_x} \frac{dq}{\ell_-^d} \textbf{1}_{q\in C_x^{\bar q}}\Big)^{n_x}
  =\frac{|C_x^{\bar q}|^{n_x}}{\ell_-^{dn_x}}
 \ee
 where: $C_x^{\bar q}= \{ r\in C_x: \text{dist}(r, \bar q_i)> R, \forall i\}$. This means that, because of 
 the presence of the hard-core, the new measure 
 ``reduces'' the admissible volume for the particles in each box. 

Let $H^{(\ell_-)}(q|\bar q)$ be
the coarse-grained Hamiltonian on scale $\ell_-$. It is
obtained by replacing
 $J^{(n)}_\ga$ by $\tilde J^{(n)}_\ga$, where
 \be\label{jcoarse}
 \tilde J^{(n)}_\ga(r_1,...,r_n) =\frac{1}{|C^{(\ell_-)}|^n} \int_{C_{r_1}^{(\ell_-)}} dq_1
 \cdots \int_{C_{r_n}^{(\ell_-)}} dq_n \, J^{(n)}_\ga(q_1,...,q_n)
 \ee
 are the coarse-grained potentials.
 
 It depends only on the particle numbers
$n_x$ (or the densities $\rho_x$) and we can thus write
        \begin{equation}
        \label{7.2.4}
 h^0( \rho  | \bar \rho  ) = H^{(\ell_-)}(q|\bar q), \qquad
 \rho_x = \rho_x^{(\ell_-)}(q),\,\, \bar \rho_x = \rho_x^{(\ell_-)}(\bar q).
        \end{equation}
Setting
        \begin{equation}
        \label{7.2.5}
\Delta H(q |\bar q )  =
H_{\ga} (q |\bar q ) - H^{(\ell_-)}(q|\bar q) 
        \end{equation}
we have  
        \begin{equation}
        \label{7.2.6}
h( \rho |\bar q) = - \sum_x \log \frac{\ell_-^{d n_x}}{ n_x !}
-\sum_{x:\, C_x^{(\ell_-)}\in\partial\La^{\text {int}}} \log Z_{x,\bar q}
+  \beta h^0( \rho  | \bar \rho  ) + \delta h( \rho |\bar q) 
      \end{equation}
where
        \begin{equation}
        \label{7.2.7}
\delta h( \rho  |\bar q) =
- \log E^0_{\rho,\bar q}\Big(    
e^{-\beta \Delta H (q |\bar q )} e^{-\beta H^{\text{hc}}_R(q)}
\sum_{\text{sp$(\Ga) \subseteq \La^0$}} 
\hat W(\Ga|q)  \Big).
        \end{equation}
It is convenient to split $\delta h( \rho|\bar q)$ in three parts
        \begin{equation}
        \label{7.2.8}
\delta h(\rho|\bar q) =  h^p(\rho|\bar q)+  h^c(\rho|\bar q)
        \end{equation}
where
        \begin{equation}
        \label{7.2.9}
 h^{p}( \rho  |\bar q) =
- \log E^0_{\rho,\bar q}\big(    
e^{-\beta \Delta H (q |\bar q )} 
e^{-\beta H^{\text{hc}}_R(q)}
  \big)
        \end{equation}
        \begin{equation}
        \label{7.2.10}
 h^c( \rho  |\bar q) =
- \log E_{\rho,\bar q}\big(    
\sum_{\text{sp$(\Ga) \subseteq \La^0$}} 
{\hat W}(\Ga|q)  \big)
        \end{equation}
        \begin{equation}
        \label{7.2.11}
 E_{\rho,\bar q} (   f ) = \frac {
 E^0_{\rho,\bar q}\big(    
e^{-\beta \Delta H (q |\bar q )} e^{-\beta H^{\text{hc}}_R(q)}
f  \big)} {E^0_\rho\big(    
e^{-\beta \Delta H (q |\bar q )}e^{-\beta H^{\text{hc}}_R(q)}\big)}.
        \end{equation}
        In words, $h^{p}(\rho|\bar q)$ is the contribution to the effective Hamiltonian coming from
        the average over the measure \eqref{newmeas} of the hard-core interaction $H^{\text{hc}}_R(q)$ and the coarse-grained 
        correction $\Delta H (q |\bar q )$ (defined in \eqref{7.2.5}). It will have an expansion in polymers as we will show in
        Section \ref{sec:clu}.
         $h^c(\rho|\bar q)$ is the same average to which is also added the contribution of the
        contours and it can also be expressed in terms of another class of polymers.

\subsection{Cluster expansion}\label{sec:clu}

In order to find an expression for $h^{p}$ and $h^c$ we perform a cluster expansion
which involves both hard-core, Kac interaction and contours.
Let us start from $h^{p}$, which is easier since there are no contours. 
We define diagrams which will be the \emph{polymers} of the cluster expansion. 
Let $L^{(2)}= (i_1,i_2)$ and $L^{(4)}=( i_1,i_2,i_3,i_4)$
denote a pair (resp. a  quadruple) of mutually distinct
particle labels. They will be called $2$-links and $4$-links. 
We will refer to the two types of $2$-links by calling them respectively $\gamma$-links and $R$-links.

\begin{definition}A  diagram  $\theta$  is a collection of $2$- and $4$-links, i.e., an ordered triple 
\be
\theta\equiv\left(\mathcal L^{(2)}_{R}(\theta), \mathcal L^{(2)}_{\ga}(\theta),
\mathcal L^{(4)}(\theta)\right),
\ee
 where 
 we denote by $\mathcal L^{(2)}_{R}(\theta)$, $\mathcal L^{(2)}_{\ga}(\theta)$ and $\mathcal L^{(4)}(\theta)$
 the set of $2$-links (of type $R$ and $\ga$) and of $4$-links in $\theta$.
Note that one can have a repetition of links, i.e., the same link $L^{(2)}$ 
can belong to both sets $\mathcal L^{(2)}_{\ga}(\theta)$ and
$\mathcal L^{(2)}_{R}(\theta)$.
We use $\mathcal L^{(2)}(\theta)$ for the set of $2$-links  
(which eventually contains twice a link when it is both a $\gamma$-link and a $R$-link)
and $\Theta$ for the set of all such diagrams.

\end{definition}

We construct the set of polymers starting from the diagrams defined above, but 
eliminating some of their links. Indeed, to work with cluster expansion an ``a priori'' estimate 
of some links is needed in order
to reduce the complexity of the diagrams that we consider. This is an essential step to assure convergence of the 
cluster expansion. To this scope, we are going to define a new set of diagrams. The procedure is the following:
We first get rid of all the $R$-links which appear over 
$\ga$-links and we extract a subdiagram $\hat\theta$.
Let $\hat\Theta\subset \Theta$ be the set of all the diagrams which do not have double $2$-links, i.e., $\hat\Theta:=\{\hat\theta: 
\mathcal L^{(2)}_{\ga}(\hat\theta)\cap \mathcal L^{(2)}_{R}(\hat\theta)=\emptyset\}$.

The next step is to obtain a diagram which is at most a tree in $R$.

\begin{definition}
(Partial ordering relation $\prec$ on a diagram $\theta$).
 For $L^{(2)}_1, L^{(2)}_2\in \mathcal L^{(2)}_{R}(\theta)$  we have that $L^{(2)}_1\prec L^{(2)}_2$
 according to lexicographic ordering (i.e., we start by comparing the first index and if the same we compare the next etc.).
 We say that a diagram is ordered if the set of its $R$-links is ordered according to this definition. We can endow an
ordered diagram with the usual notion of distance. We will write $d(v)$ to indicate the distance of a vertex $v$ to the 
first vertex in the previous order relation.
\end{definition}

\begin{definition}\label{def1}
(Redundant link).
 Given an ordered diagram $\theta$, we say that a link $L^{(2)}\in
 \LLd_{R}(\theta)$ is redundant in the following two cases:
\begin{itemize}
\item If $L^{(2)}=\{i,j\}$ with $d(i)=d(j)$;
\item If $L_1^{(2)}=\{i_1,j\}$ with $d(i_1)=d(j)-1$ and it exists
  $L_2^{(2)}=\{i_2,j\} \in \LLd_{R}(\theta)$, with $d(i_2)=d(j)-1$, such
  that: $L_2^{(2)} \prec L_1^{(2)}$ (i.e., $i_2<i_1$). 
\end{itemize}
We denote the set of the redundant links of a diagram $\theta$ by: $\RRd_{R}(\theta)$. 
\end{definition}

We call $\bar\Theta\subset \hat\Theta$ the set of diagrams with no double $2$-links and with no redundant links.
In formulas: $\bar\Theta:=\{\bar\theta: \bar\theta\in \hat\Theta, \RRd_{R}(\bar\theta)=\emptyset\}$. 

 Two diagrams $\theta$ and $\theta'$
are  \emph{compatible} ($\theta\sim\theta'$) if the set of their common labels 
is empty.

\begin{theorem}\label{one}
For all $\ga$ and $R$ small enough,
there exist functions $z_{\ga,R}^T(\pi;\rho; \bar q)$ such that
        \begin{equation}
        \label{7.2.26}
 h^{p}( \rho  |\bar q) =
- \sum_{\pi} z_{\ga,R}^T(\pi;\rho;\bar q),
        \end{equation}
where $\pi$ is a collection of non-compatible diagrams in the space $\bar\Theta$.
\end{theorem}

Let us now find an expansion for $h^c$ defined in \eqref{7.2.10}.
Let us fix a collection $\underline\Ga= \{\Ga_i\}_{i=1}^n$,
where $\Ga_i\equiv (\text{sp}(\Ga_i),\eta_{\Ga_i})$. As said after \eqref{9.1a.4.1}, 
the weights  $W^{\pm}_{\ga,R,\la}(\Ga_i; q^{\pm})$ depend only
on $q^{D_i}$, i.e., the restriction of $q^{\pm}$ to $D_i=\{r\in c(\Ga)^c: {\rm
dist}(r,c(\Ga))\le 2\ga^{-1}\}$.  We also let $D:=\cup_{i=1}^n D_i$.
We then have, for the numerator of \eqref{7.2.10},
\be
 E^0_{\rho^D,\bar q}\Big( W(\Ga|q)e^{-h^{p}(\rho^{\La\setminus D}|\bar q\cup q^D)}\Big).
\ee
We write $h^{p}$ as a sum of
 clusters using  \eqref{7.2.26}. 
 Due to the dependence of the a priori measure on $\bar q$
 (now on both $\bar q$ and $q^D$),
 the clusters involving a particle in a neighboring $\ell_-$-cell to $D$
 will also depend on $q^D$.
 We denote the union of the set $D$ with the frame
 consisting of the neighboring $\ell_-$-cells by $D^*\in\mathcal D^{(\ell_2)}$.

 To distinguish between 
 clusters we introduce 
 $\bar D_i:=D_i\cup\{r: \text{dist}(r,D_i) \le \ell_+/4\}\in\mathcal D^{(\ell_-)}$
and we call
 $\mathcal B_i$ the set of all
clusters  $\pi$ whose points are all in $\bar D_i$.
As the distance between contours is $\ge \ell_+$, the
sets  $\mathcal B_i$ are mutually disjoint; we call
 $\mathcal B$ their union. Note that they depend on 
$\underline\Ga$
through the domain where they are constructed.
By
$\mathcal R_i$ we denote the set of $\pi$ which have 
points both in $D_i^*$ (so that they depend on $q^D$) and 
 in the complement of $\bar D_i$
(such $\pi$ are therefore
not in  $\mathcal B_i$). There may be
points of  $\pi \in \mathcal R_i$ which are
 in  $D_j^*$, $j\ne i$, hence also $\pi \in \mathcal R_j$,
 so that
the sets  $\mathcal R_i$ are not disjoint.  We call
 $\mathcal R$ their union.  

For any given $\underline\Ga$ we  do  analogous splitting
on the polymers appearing when developing the denominator
of \eqref{7.2.11} thus defining the sets
$\mathcal B'_i$, $\mathcal B'$,
$\mathcal R'_i$, $\mathcal R'$.
The clusters that appear in the numerator and denominator
of \eqref{7.2.11} are different, however those not
in $\mathcal B\cup \mathcal R$ 
(i.e., those that do not involve $q^D$)
are common to the corresponding ones in the denominator of \eqref{7.2.11} 
(i.e., those not
in $\mathcal B'\cup \mathcal R'$) and have same
statistical weights, hence they cancel. 

The clusters  $\pi\in\mathcal B$
can be grouped together and absorbed by a renormalization of the measure in 
$E^0_{\rho^D,\bar q} $,
since they do not involve interactions between different contours.
Thus, they will be part of the  \emph{activities} in the expansion, while the
\emph{polymers} will be defined in terms of elements of $\mathcal R$ and $\mathcal R'$.

Hence, to formulate the problem into the general context of the
abstract polymer model
we define as {\it connected} polymer $P$ a set of contours with
``connections" consisting of elements of $\mathcal R\cup\mathcal R'$ 
which necessarily ``connect" all
contours in the given set and ``decorations" consisting of clusters in
$\mathcal R\cup\mathcal R'$
not necessarily connecting contours. We denote by $\mathcal P$ the space of
all such elements
\begin{eqnarray}\label{}
\mathcal P & := &\left\{
P\equiv (\underline\Ga(P),R(P)), \forall \Ga_i,\Ga_j \in \underline \Ga(P), \exists
\pi \in R(P)\subset\mathcal R\cup\mathcal R'
\right.\nonumber\\
&&\left.
\text{connecting }\, D_i^*,D_j^*\in D^*(\underline\Ga)
\right\}.
\end{eqnarray}
We use $D(P), D^*(P)$ to denote the set of 
frames corresponding to the 
contours in $P$ and $R(P)$ to denote the set of clusters.
We also introduce $A(\pi)$ to denote 
the union of the $C^{(\ell_2)}$ cells
which correspond to the labels of $\pi$.
Similarly, let $A(P):=\cup_{\Ga\in\underline\Ga(P)}D^*(\Ga)\cup_{\pi\in R(P)}A(\pi)$.
A compatible collection of polymers
 consists of mutually compatible
polymers.

\begin{theorem}\label{two}
For all $\ga$ and $R$ small enough,
there exist functions $\zeta_{\ga,R}^T(C;\rho)$ such that
  \begin{equation}
        \label{7.2.41}
 h^c( \rho  |\bar q) =-\sum_{C} \zeta_{\ga,R}^T(C;\rho),
          \end{equation}
where $C$ is a collection of non-compatible polymers $P$ in the space $\mathcal P$.
\end{theorem}
With these two theorems we can define a new measure on the space of the density configurations $\rho=\{\rho_x\}_{x\in X^{(\ell_-)}_{\La}}$, where the new Hamiltonian is

 \begin{equation}\label{effham}
h( \rho |\bar q) = - \sum_x \log \frac{\ell_2^{d n_x}}{ n_x !}
-\sum_{x:\, C_x^{(\ell_2)}\in\partial\La^{\text {int}}} \log Z_{x,\bar q}
+  \beta h^0( \rho  | \bar \rho  ) +  h^{p}( \rho  |\bar q) +h^c( \rho  |\bar q). 
      \end{equation}

We then use the notation $\mathbb E$ for the expectation w.r.t. this new coarse-grained measure $\nu$.

To estimate the difference $ E_{\mu^1} \big(f_{x_1,..,x_n} \big)-E_{\mu^2} \big(f_{x_1,..,x_n} \big)$ in Theorem \ref{surcorr},
we split  $ E_{\mu^i} \big(f_{x_1,..,x_n} \big)$ into two parts, one which is of order one and one which is exponentially small. However
the order one parts will be small when we consider their difference. 
The main idea is the following. 

Given $x_1,...,x_n$, for $n=1,2,4$, and such that each $x_i,x_j$ are not more distant than $\ga^{-1}$, we choose a box of side $2\ell_+$
that contains all of them and is far enough from the boundary.  
The contribution of clusters attached to any subset of $x_1,..., x_n$ inside the box will be denoted by $g$ and the ones attached to any subset of $x_1,..., x_n$ inside the box and going out of it will be denoted by $R$.
This latter contribution is exponentially small, as a corollary of the above theorems. 

We first prove this splitting in the following Lemma:

\begin{lemma} \label{lemma1.4.1}
Let $f_{x_1,..,x_n}$ be as in \eqref{expec}, then
      \begin{equation}
      \label{7.3.1}
  E_{\mu^i} \big(f_{x_1,..,x_n} \big)
=  \mathbb E_{\nu^i} (g)+ R_i,\qquad i=1,2
      \end{equation}
where  
$g$ is a function of
$\{\rho_x\}$   with $x \in X_\La$ contained
in the cube of side $2\ell_+$ and $R_i$
are remainder terms.  
Moreover,
  there are $\delta>0$
and constants $c_1,c_2,c$ so that
      \begin{equation}
      \label{7.3.2}
\|g\|_{\infty} \le c_1,\qquad 
\|R_i\|_{\infty} \le 
c_2
 e^{-c\ga^{-\delta}}.
      \end{equation}
 
\end{lemma}

To conclude the proof of Theorem \ref{surcorr} we need to estimate the difference $\mathbb E_1(g)-\mathbb E_2(g)$. 
In order to do this, we work in  the coarse-grained model, i.e., in the space
   \be\label{space}
\mathcal X^\La=\Big \{
\und n = (n_x)_{x\in X_\La}\in \mathbb N^{X_\La}:|\ell_-^{-d}
n_x-\rho_{\beta,+}|\le
\zeta,
\;
\text{for all $x\in X_\La$}\Big\}, \quad n_x=\ell_-\rho_x.
          \ee
The goal is to prove that there exists a joint representation $\mathcal P(\und n^1,
\und n^2|{\bar q}^1,{\bar q}^2)$ 
of the measures $\nu^1$ and $\nu^2$ on $\mathcal X^{\La}$ 
such that, for any $x\in X_\La$, and denoting by $\mathcal E$ the expectation
w.r.t.\ to $\mathcal P$, we can bound the difference $\mathbb E_{\nu^1}(g)-\mathbb E_{\nu^2}(g)$ with
$\mathcal E \big[ d(n_x^1,n_x^2)\big]$, where $d(n_x^1,n_x^2)$ is an appropriate distance that we have to
define and where $\mathcal E \big[ d(n_x^1,n_x^2)\big]$ has the desired exponential decay property.

To complete the proof we then need to find a bound for $\mathcal E \big[ d(n_x^1,n_x^2)\big]$. This 
comes from a Dobrushin uniqueness condition. We want to bound the Vaserstein distance
between
two Gibbs measures with the same Hamiltonian
\eqref{effham} but with
different b.c.\ ${\bar q}^i$, $i=1,2$.  It is convenient
to define the Vaserstein distance in terms of the following cost functions
        \begin{equation}
        \label{7.4.5}
d(\und n^1,\und n^2)= \sum_{x\in X_\La} d(n_x^1, n_x^2),\qquad
d(n_x^1, n_x^2) = |n_x^1-\und n_x^2|
          \end{equation}
          and if we suppose $\bar q^1=(\bar q^1_1,...,\bar q^1_n)$ and $\bar q^2=(\bar q^2_1,...,\bar q^2_{n+p})$,
\be\label{discr}
D_z({\bar q}^1,{\bar q}^2):= 
p + \min_{\{j_{\ell}\}} \sum_{\ell=1}^n \mathbf{1}_{\bar q^1_{\ell} \neq \bar q^2_{j_{\ell}}},
 \ee
 the $\min$ being over all the subsets $\{j_{\ell}\}$ of $\{1,...,n+p\}$ which have cardinality $n$.
Following Dobrushin, we need to estimate
the Vaserstein distance between conditional  probabilities at
a single site. We thus fix arbitrarily $x\in \La$,
$\und n^i$, $i=1,2$,  in $\mathcal X^{
\La\setminus x}$, call $\rho^i:=\ell_2^{-d}\und n^i$;
  $\bar q^i$ are the b.c.\ 
outside $\La$.  
The energy in $x$ plus the interaction with the
outside is, as usual, 
        \begin{equation}
        \label{7.4.6}
h(\rho_x|\rho^i,\bar q^i) =  h\big(\{\rho_x,\rho^i\}|\bar q^i\big)
- h(\rho^i|\bar q^i),
          \end{equation}
where the first term on the r.h.s.\
is the energy of the configuration
$\{\rho_x,\rho^i\}$ (with $\bar q^i$ outside $\La$).
The second term is the energy in  $\La\setminus x$
of $\rho^i$ with nothing in $x$ and
$\bar q^i$ outside $\La$.   
The conditional Gibbs measures  are then the following probabilities
on $\mathcal X^{x}$ (for $i=1,2$)
        \begin{equation}
        \label{7.4.8}
p(n_x|\rho^i,\bar q^i) = \frac 1{Z_x(\rho^i,\bar q^i)}
\exp\big\{-
h(\rho_x|\rho^i,\bar q^i)\big\},
          \end{equation}
and their Vaserstein distance is
        \begin{equation}
        \label{7.4.9}
R\Big( p(\cdot|\rho^1,\bar q^1),p(\cdot|\rho^2,\bar q^2)\Big)
 := \inf_{Q} \sum_{n_x^1,n_x^2} Q(n_x^1,n_x^2)d(n_x^1,n_x^2),
          \end{equation}
where the inf is over all the joint representations $Q$
of
$p(\rho_x|\rho^i,\bar q^i)$, $i=1,2$.  The key bound for
the Dobrushin scheme to work is the following theorem.
\begin{theorem}
There are $u, c_1,c_2 >0$  s.t. $\forall x \in \La$
\[
R\Big( p^1(\cdot),p^2(\cdot)\Big)
 \le \sum_{z\in X_\La, z\ne x} r_{\ga,R}(x,z) d(n_z^1,n_z^2) 
+ \sum_{z\in X_{\La^c}} r_{\ga,R}(x,z) D_z({\bar q}^1,{\bar q}^2) 
\]
with
    \[
\sum_{z} r_{\ga,R}(x,z) \le u <1
          \]
        \[
 r_{\ga,R}(x,z) \le c_1 e^{-c_2\ga |z-x|}, \quad |z-x|\ge \ell_+.
          \]
\end{theorem}

\vspace{0.5cm}

\begin{remark}
{\rm
The reduction to an abstract contour model allows us
to deal with a coarse-grained system in which the
configurations we look at are those chosen in the restricted ensembles, roughly speaking
those which should be seen under the effects of a double-well potential once we restrict to
its minima. In this scenario, after we compute the effective hamiltonian for the coarse-grained 
system, we have a new Gibbs measure which depends only on the cells variables. But now these variables 
are close to the mean-field value and in such a setup it is possible to prove the validity of the Dobrushin
uniqueness theory.}
\end{remark}

%


\end{document}